\documentclass[a4paper,UKenglish]{lipics}

\usepackage{microtype}
\usepackage{graphicx}
\usepackage{amsmath}
\usepackage{amssymb}
\usepackage{enumerate}
\usepackage{varioref}
\usepackage{float}
\usepackage{url}
\newcommand\nc{\newcommand}

\newtheorem{rules}{Rule}

\nc{\crl}[2]{\begin{corollary}\label{crl:#1} #2 \end{corollary}}
\nc{\dfn}[2]{\begin{definition}\label{def:#1} #2 \end{definition}}
\nc{\lem}[2]{\begin{lemma}\label{lem:#1} #2 \end{lemma}}
\nc{\prp}[2]{\begin{proposition}\label{prp:#1} #2
\end{proposition}}
\nc{\thm}[2]{\begin{theorem}\label{thm:#1} #2\end{theorem}}
\nc{\fac}[2]{\begin{lemma}\label{fact:#1} #2 \end{lemma}}
\nc{\rul}[2]{\begin{rules}\label{rul:#1} #2 \end{rules}}

\nc{\eqn}[2]{\begin{eqnarray}\label{eqn:#1} #2 \end{eqnarray}}
\nc{\tbl}[3]{\begin{table}[hbt] #3 \caption{#2} \label{tab:#1}
\end{table}}
\nc{\refc}[1]{Corollary~\ref{crl:#1}}
\nc{\refd}[1]{Definition~\ref{def:#1}}
\nc{\refl}[1]{Lemma~\ref{lem:#1}}
\nc{\refp}[1]{Proposition~\ref{prp:#1}}
\nc{\reft}[1]{Theorem~\ref{thm:#1}} \nc{\refe}[1]{(\ref{eqn:#1})}
\nc{\reftb}[1]{Table~\ref{tab:#1}}
\nc{\reffc}[1]{Fact~\ref{fact:#1}}
\nc{\refr}[1]{Rule~\ref{rul:#1}}
\nc{\pf}[1]{ \noindent \emph{Proof.} #1 \hfill \qed\par}

\long\def\invis#1{}

\title{Linear Kernels for Separating a Graph into Components of Bounded Size\footnote{Supported by NFSC of China under the Grant 61370071.}}
\titlerunning{Kernels for Separating a Graph into Components of Bounded Size} 

\author[1]{Mingyu Xiao}
\affil[1]{School of Computer Science and Engineering\\
University of Electronic Science and Technology of China,China\\
  \texttt{myxiao@gmail.com}}
\authorrunning{M. Xiao} 

\Copyright{M. Xiao}

\subjclass{Dummy classification -- please refer to \url{http://www.acm.org/about/class/ccs98-html}}
\keywords{Linear Kernels; Graph Algorithms; FPT; Balanced Separators}

\serieslogo{}
\volumeinfo
  {Billy Editor and Bill Editors}
  {2}
  {Conference title on which this volume is based on}
  {1}
  {1}
  {1}
\EventShortName{}
\DOI{10.4230/LIPIcs.xxx.yyy.p}

\begin{document}

\maketitle

\begin{abstract}
Graph separation and partitioning are fundamental problems that
have been extensively studied both in theory and practice.
The \textsc{$p$-Size Separator} problem, closely related to the \textsc{Balanced Separator} problem, is to check whether we can delete at most $k$ vertices in a given graph $G$ such that each connected component of the remaining graph has at most $p$ vertices.
This problem is NP-hard for each fixed integer $p\geq 1$ and it becomes the famous \textsc{Vertex Cover} problem when $p=1$.
It is known that the problem with parameter $k$ is W[1]-hard for unfixed $p$. 
In this paper, we prove a kernel of $O(pk)$ vertices for this problem, i.e., a linear vertex kernel for each fixed $p \geq 1$.
In fact, we first obtain an $O(p^2k)$ vertex kernel by using a nontrivial extension of the expansion lemma.
Then we further reduce the kernel size to $O(pk)$ by using some `local adjustment' techniques.
Our proofs are based on extremal combinatorial arguments and
the main result can be regarded as a generalization of the  Nemhauser and Trotter's theorem for the \textsc{Vertex Cover} problem.
These techniques are possible to be used to improve kernel sizes for more problems, especially
problems with kernelization algorithms based on techniques similar to the expansion lemma or crown decompositions.

 \end{abstract}
\section{Introduction}

Finding optimal separators and cuts in graphs is a classical topic in combinatorial optimization.
Different versions of graph separation and graph cut problems have been studied extensively from the perspective of approximation algorithms and heuristics. In recent years there has
been an increase of interest in parameterized algorithms of such problems~\cite{BDT:multicut,dfvs,bisection,feige:separator,kt:kway,Marx:separation,Marx:separator,Marx:multicut,H:separator2,2sat,xiao:mcut}.

Graph separators play a mysterious and not yet fully understood
role in parameterized algorithms of certain problems. Proving that \textsc{Bipartization}~\cite{oddcycle}, \textsc{Multicut}~\cite{BDT:multicut,Marx:multicut}, \textsc{$k$-Way Cut}~\cite{kt:kway},
\textsc{Directed Feedback Vertex Set}~\cite{dfvs}, \textsc{Almost $2$-Sat}~\cite{2sat} and \textsc{Minimum Bisection}~\cite{bisection} are fixed-parameter tractable (FPT) answered longstanding open
questions in parameterized complexity, and in each case the algorithm relies on a non-obvious use of separators.

Balanced separator is one of the most important topics~\cite{feige:separator0,feige:bisection,feige:separator,Marx:separation}. The \textsc{Balanced Separator} problem is to check
whether there is a vertex separator of size $k$
that partitions a given $n$-vertex graph into connected components of size at most $\alpha n$ $(0< \alpha <1)$. This vertex separator is also called an \emph{$\alpha$-separator}.
\textsc{Balanced Separator} can be described in another way, which is called the \textsc{$p$-Size Separator} problem.
A \emph{$p$-size separator} of an $n$-vertex graph is a vertex separator of size $k$
that partitions the graph into connected components of size at most $p$. We can see that a $p$-size separator is
 an $\alpha$-separator and vice versa if $p=\alpha n$.

Balanced separators are used in applications of designing divide-and-conquer algorithms and parallel algorithms
for a large number of graph problems. Enright and Meeks~\cite{EM} addressed some applications of these separators from real life, such as to restrict the size of an epidemic.
\textsc{Minimum Bisection}~\cite{bisection} is also equivalent to finding a balanced edge-separator.
There are several approximation algorithms to find small balanced separators~\cite{feige:separator0,feige:bisection,feige:separator,BMN}.
In terms of parameterized algorithms,
the problem with parameter $k$ is W[1]-hard for unbounded size of the components (i.e., $p$ or $\alpha n$ is not a constant)~\cite{Marx:separation,feige:separator}.
A trivial branch-and-reduce algorithm that tries all $p+1$ possibilities for a connected subgraph of $p+1$ vertices
gets the running time bound of $O^*((p+1)^k)$, which implies that the problem with  parameter $k$ is FPT for each fixed $p$.
In terms of kernelizations, a kernel of $O(p^2k^2)$ vertices is known for \textsc{$p$-Size Separator}~\cite{ddh}.
This is a quadratic vertex kernel for each fixed $p$. This paper will give the first linear vertex kernel for this problem.
Our main result is \reft{our-thm}, which can be regarded as a generalization of the  Nemhauser and Trotter's local optimization theorem for \textsc{Vertex Cover}, and we use only extremal combinatorial arguments to prove it.

When $p=1$, \textsc{$p$-Size Separator} becomes \textsc{Vertex Cover}.
Nemhauser and Trotter~\cite{NT-theorem}
proved a famous theorem (NT-Theorem) for  \textsc{Vertex Cover}.
\thm{ntthm}{ \emph{[\textbf{NT-Theorem}]}
For an undirected graph $G=(V,E)$ of $n=|V|$ vertices and $m=|E|$ edges, there is an $O(\sqrt{n}m)$-time
algorithm to compute two disjoint vertex subsets $C$
and $I$ of $G$ such that for any minimum vertex cover $K$ of the induced subgraph $G[V \setminus (C \cup I)]$, $K \cup C$ is
a minimum vertex cover of $G$ and
$$|V \setminus (C \cup I)|\leq 2|K|.$$
}
This theorem was proved by constructing an algorithm based on linear programming relaxation~\cite{NT-theorem} and
the algorithm can be used to reduce the size of the input
graph by possibly finding partial solution.
The NT-Theorem has great applications in approximation algorithms~\cite{BE:WVC,hochbaum,khuller} and parameterized algorithms~\cite{CKJ:VC,A:crown2}. We can see that $V\setminus I$ is a 2-approximation solution
and $G[V \setminus (C \cup I)]$ is a $2k$-vertex kernel
with $k$ being the size of the vertex cover.
Due to NT-Theorem's practical usefulness and theoretical depth in graph theory, it has attracted numerous
further studies and follow-up work~\cite{FG:gNT,BRH:extension,CC:WVC,A:crown2,X:gNT}.
In this paper, we will prove the following local optimization theorem similar to NT-Theorem for \textsc{$p$-Size Separator}.

\thm{our-thm}{ \emph{\textbf{[Our result]}}
For an undirected graph $G=(V,E)$ of $n=|V|$ vertices and $m=|E|$ edges, there is an $O(n^{4}m)$-time
algorithm to compute two disjoint vertex subsets $C$
and $I$ of $G$ such that for any minimum $p$-size separator $K$ of the induced subgraph $G[V \setminus (C \cup I)]$, $K \cup C$ is
a minimum $p$-size separator of $G$ and
\[|V \setminus (C \cup I)|\leq 9p |K|. \]
}

\reft{our-thm} implies a kernel of $9pk$ vertices for \textsc{$p$-Size Separator}
with $k$ being the size of the solution,
which is linear in $k$ for any constant $p\geq 1$.
Dell and van Melkebeek~\cite{dell} proved that \textsc{Vertex Cover} and some related problems
do not have kernels consisting of $O(k^{2-\epsilon})$ edges for any constant $\epsilon>0$
unless the polynomial-time hierarchy collapses.
This also implies that linear size would be the best possible bound on the number of vertices in any kernel of \textsc{$p$-Size Separator} for each fixed $p\geq 1$. It is also known that
\textsc{$p$-Size Separator} with parameter $k$ is W[1]-hard for unbounded $p$~\cite{Marx:separation,feige:separator}.
Then it is unlikely to remove $p$ from the size function of any kernel for this problem.
These two hardness results also imply  that our kernel result is somewhat close to optimal.

To prove \reft{our-thm}, we will use a decomposition, called `weighted crown decomposition'.
It can be regarded as an extension of the crown decomposition for \textsc{Vertex Cover}~\cite{A:crown,cfj:crown}
 as well the structure under the expansion lemma (Lemma 8 in~\cite{fomin}).
 To find a weighted crown decomposition, we need to consider a bipartite graph and find vertex-disjoint stars in it
 with leaves from one side and centers from another side (also having some other properties). This technique has been used to obtain kernels in
 many problems~\cite{fvs,FG:gNT,fomin,X:gNT}. The expansion lemma~\cite{fomin} provides a condition for the existence of such star packings in bipartite graphs. For our problem and the weighted crown decomposition, we need to
 consider a vertex-weighted bipartite graph and find stars in it.
 So we prove a lemma (\refl{wexpansion}) for the weighted case. We call this lemma the weighted expansion lemma for convenience. We would like to mention that when the graph is a weighted graph, it will become hard to compute such
 star packings and we may need to relax the size of the bipartite graph to obtain polynomial time algorithms.
 So our algorithm for weighted expansion lemma is different from that for normal expansion lemma,
 not just a simple extension.
 By using the weighted expansion lemma, we can obtain a kernel of size $O(p^2k)$.
 A more interesting part should be the improvement from $O(p^2k)$ to $O(pk)$.
 We will use some techniques to iteratively adjust some local structures.
  Finally, we can apply the weighted expansion lemma to only a (small) part of
 the original bipartite graph and then get the bound of $O(pk)$.

 Some figures and proofs are moved to Appendix due to space limitation. 


\section{Preliminaries}\label{sec_pre}
Let $G=(V,E)$ denote a graph with vertex set $V$ and edge set $E$. We will use $n$ and $m$ to denote
the number of vertices and edges of our input graph $G$, respectively.
We also use $\gamma_p (G)$ to denote the size of a minimum $p$-size separator of $G$ for any integer $p\geq 1$.
For a vertex subset $V'\subseteq V$,
the subgraph induced on $V'$ is denoted by $G[V']$ and $G[V\setminus V']$ may be written as $G\setminus V'$.
For a graph $G'$, we use $V(G')$ and $E(G')$ to denote the vertex set and edge set of $G'$, respectively.
We say that a vertex set $V'$ (resp., a subgraph $G'$) is \emph{adjacent} to a vertex $v\not\in V'$ (resp., $v\not\in V(G')$) if there is at least one edge between $v$ and a vertex in $V'$ (resp., $G'$).
For a vertex (resp., edge) subset $V_0$, a vertex (resp., edge) in $V_0$ is called a \emph{$V_0$-vertex} (resp., \emph{$V_0$-edge}).
A bipartite graph with edges between two vertex sets $V_1$ and $V_2$ is denoted by $(V_1,V_2, E_{12})$, where
$E_{12}$ is the set of edges between $V_1$ and $V_2$. A set $\{v\}$ of a single element may be simply written as $v$.

For an integer $d\geq 1$, a connected subgraph with $d$ vertices is called a \emph{$d$-subgraph}.
A set $\mathcal{S}$ of vertex-disjoint $d$-subgraphs is called a \emph{$d$-subgraph packing}.
A star with exactly $d+1$ vertices is called a  {\em $d$-star}.
The unique vertex of degree $>1$ in a $d$-star with $d> 1$ is called the \emph{center}
of the star and all other degree-1 vertices are called the \emph{leaves} of the star.
For a 1-star, any vertex can be regarded as a \emph{center} and the other vertex as a \emph{leaf}.
A star with a center $v$ is also called a star \emph{centered at} $v$.
For two disjoint vertex sets $V_1$ and $V_2$, a star with the center in $V_1$ and all leaves in $V_2$ is called a \emph{star from $V_1$ to $V_2$}. A \emph{full star packing from $V_1$ to $V_2$} is a star packing of size equal to $|V_1|$.
A star with vertex-weighted is called a \emph{$(p_1,p_2)$-star} if
the total vertex weight in the star is at least $p_1$ and at most $p_2$.
In this paper, we will always set the vertex weight zero for the center of a star. So a weight $(p_1,p_2)$-star will mean a star with total leaf weight
at least $p_1$ and at most $p_2$.


\section{The weighted expansion lemma and an $O(p^2k)$ kernel} \label{kernel1}
We give an algorithm that returns a kernel of $O(p^2k)$ vertices for \textsc{$p$-Size Separator}.
We mainly use the weighted expansion lemma (\refl{wexpansion}) to obtain this result.

For a graph $G=(V,E)$ with a partition $(A,B=V\setminus A)$, we define the \emph{auxiliary bipartite graph} $H_G=(A',B',E')$ of $G$  as follows:
each vertex $a\in A'$ is corresponding to a connected component $C_a$ in the induced graph $G[A]$; the weight $w(a)$ of $a\in A'$ is the number of vertices in $C_a$;
$B'$ is a copy of $B$, the weight $w(b)$ of each $b\in B'$ is $0$, and a vertex $a\in A'$ is adjacent to a vertex $b\in B'$ if and only if
there is a vertex in $C_a$ adjacent to $b\in B$ in the graph $G$.
Note that the auxiliary bipartite graph is a vertex-weighted graph.
Please refer to Figure~\ref{figt1} in Appendix~\ref{fig} for an illustration for auxiliary bipartite graphs.
This graph will be frequently used in our algorithms and analysis.

\dfn{weight_dc}{[\textbf{Weighted Crown Decomposition}]
 For an integer $p\geq1$, a $p$-weighted crown decomposition of a graph $G=(V,E)$ is
a partition of the vertex set of $G$ into three sets $I$, $C$ and $J$ such that\\
(1) there are no edges between $I$ and $J$, \\
(2) each connected component in the induced graph $G[I]$ has at most $p$ vertices, and \\
(3) there is a full $(p,\infty)$-star packing from $C'$ to $I'$ in the auxiliary bipartite graph $H_G=(I',C',E')$ of $G[I\cup C]$ with partition $(I,C)$.
}

Figure~\ref{figt2} in Appendix~\ref{fig} gives an illustration for weighted crown decompositions and normal crown decompositions.


\lem{wcrown}{Let $(I,C,J)$ be a $p$-weighted crown decomposition of a graph $G=(V,E)$. Then $(I,C)$ satisfies the local optimality condition in \reft{our-thm}, i.e.,
$K \cup C$ is a minimum $p$-size separator of $G$ for any minimum $p$-size separator $K$ of the induced subgraph $G[J]$.
}

A proof for this lemma can be found in Appendix~\ref{ap_proof}.
Based on \refl{wcrown}, we can switch to find weighted crown decompositions instead of computing
the sets $I$ and $C$ in \reft{our-thm} directly.
Here arises a question how to find weighted crown decompositions of a graph?


\lem{wexpansion}{
\emph{\textbf{[Weighted Expansion Lemma]}}
Let $p\geq 1$ be an integer and $(A,B=V\setminus A)$ be a partition of the vertex set of a graph $G=(V,E)$
such that each connected component of $G[A]$ has at most $p$ vertices.
There are sets $I \subseteq A$ and $C \subseteq B$ such that $(I,C,J=V\setminus(I\cup C))$ is a
$p$-weighted crown decomposition of $G$ and $(I,C)$ satisfies the size condition:
$$|A\setminus I|\leq (2p-1)|B\setminus C|.$$
Furthermore, $(I,C)$ can be computed in $O(|V|^2|E|)$ time.
}


We mainly use an algorithm ${\tt crown}(G,B,p)$ to prove \refl{wexpansion}.
The algorithm ${\tt crown}(G,B,p)$ takes a graph $G=(V,E)$, a vertex subset $B\subseteq V$ and an integer $p\geq 1$ as the input, and outputs two sets $I \subseteq V\setminus B$ and $C \subseteq B$ such that $(I,C,J=V\setminus(I\cup C))$ is a
$p$-weighted crown decomposition of $G$.
The first step of ${\tt crown}(G,B,p)$ is to compute the auxiliary bipartite graph $H_G=(A',B',E')$ of $G$ with partition $(V\setminus B,B)$.
In the following steps, the algorithm mainly deals with the graph $H_G$ and always maintains an edge set $M$ of $H_G$ such that each vertex in $A'$ has exactly one edge in $M$ (then the edges in $M$
will form a star packing from $B'$ to $A'$).
Such an edge set $M$ is obtained initially by, for each vertex $v\in A'$, selecting an arbitrary edge incident on $v$ to $M$. It is done in Step~2.
In the next steps, we may modify $M$ by replacing an $M$-edge incident on a vertex $v\in A'$ with another edge incident on $v\in A'$. This operation will not destroy the above property of $M$.

For a fixed $M$, we define the following notations.
A path $P$ in $H_G$ that alternates between edges in $M$ and edges not in $M$ is called an \emph{$M$-alternating path}.
An $M$-alternating path is called a  \emph{strong $M$-alternating path} from a vertex $v$ to $u$ if
 the first edge (the edge incident on $v$) in $P$ is in $M$.
The edges in $M$ form a star packing $Q$ from $B'$ to $A'$. We partition $Q$ into three subsets $Q_1, Q_2$ and $Q_3$, where $Q_1$ is the set of stars with total vertex weight at least $2p$,
$Q_2$ is the set of stars with total vertex weight at least $p$ but less than $2p$, and $Q_3$ is the set of stars with total vertex weight less than $p$.
Let $A'_i=A'\cap V(Q_i)$ and  $B'_i=B'\cap V(Q_i)$ for $i=1,2,3$.

There is a full $(p,\infty)$-star packing from $B'_1\cup B'_2$ to $A'_1\cup A'_2$ in $H_G$. It is possible to use
$B'_1\cup B'_2$ and $A'_1\cup A'_2$  to construct $C$ and $I$ in \refl{wexpansion}.
In fact, if there is no strong $M$-alternating path from a vertex in $B'_1$ to a vertex in $B'_3$ then
we can find a subset of $A'_1\cup A'_2$ and a subset of  $B'_1\cup B'_2$  to identify a solution $(I,C)$ (this will be proved later).
Our idea is to destroy $M$-alternating paths from vertices in $B'_1$ to vertices in $B'_3$ by
iteratively replacing an $M$-edge incident on a vertex $v\in A'$ with another edge incident on $v\in A'$.
One may think that if there is an $M$-alternating path $P$ from a $B'_1$-vertex to a $B'_3$-vertex, we can
replace $M\cap E(P)$ by $E(P)\setminus M$ in $M$ to destroy $P$ as what we do for unweighted cases.
However, this replacement may create new $B'_1$-vertices and cause some trouble in the analysis.
To avoid possible endless loops, we need to use a `hierarchical structure' of $H_G$.

In the \emph{hierarchical structure}, we classify some vertices in $H_G$ into different levels based on $M$:\\
(i) all vertices in $B'_1$ are in level-$0$;\\
(ii) for each odd $i\geq 1$, a vertex is in level-$i$ if and only if it is adjacent to a vertex in level-$(i-1)$ via an $M$-edge;\\
(iii) for each even $i\geq 1$, a vertex is in level-$i$ if and only if it has not be assigned to a level before and it is adjacent to a vertex in level-$(i-1)$ via an edge not in $M$.

\begin{figure}[h]
\begin{center}
\includegraphics[width=1\textwidth]{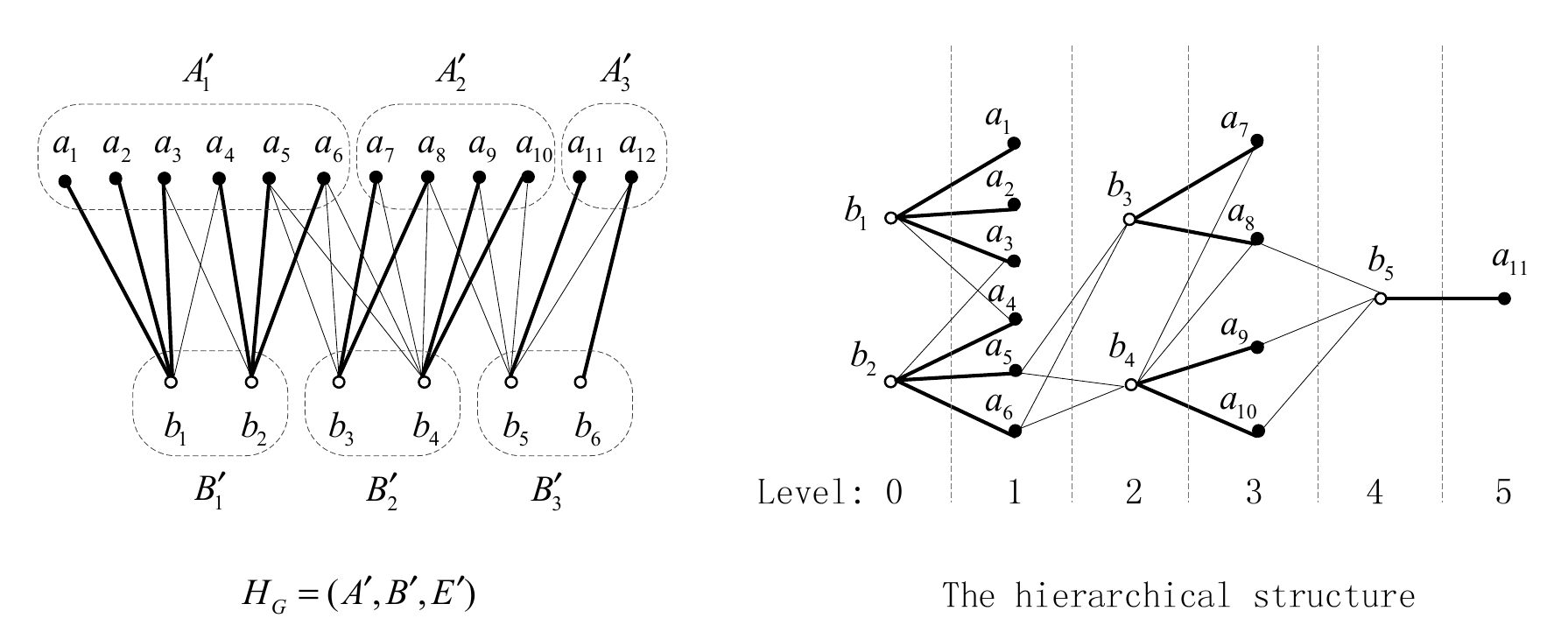}
\end{center}
\caption{Graph $H_G$ and the hierarchical structure}\label{figt3}
\end{figure}

The hierarchical structure can be built in linear time via a breadth-first search (BFS).
Figure~\ref{figt3} gives an illustration for the hierarchical structure.
Note that some vertices may not be included in the hierarchical structure.
These vertices can be simply ignored in our algorithm.

A strong $M$-alternative path
containing exactly $i+1$ vertices from a vertex $v_0$ in level-0 to a vertex $u$ in level-$i$ is called a
\emph{pure $M$-alternative path} from $v_0$ to $u$.
It is easy to see the following properties in a hierarchical structure:\\
(\textbf{P1}) for each even number $i$ all vertices in level-$i$ are $B'$-vertices, and
for each odd number $i$ all vertices in level-$i$ are $A'$-vertices;\\
(\textbf{P2})
a vertex appears in the hierarchical structure if and only if there is a strong
$M$-alternative path
 from a vertex in level-0 to it;\\
(\textbf{P3})
for any vertex in the hierarchical structure, there is a pure $M$-alternative path
 from a vertex in level-0 to it.

A vertex $v$ in a hierarchical structure is called a \emph{redundant vertex} if \\
(i) it is in level-$i$ with an odd number $i$, and \\
(ii) there is a neighbor $u$ of $v$ in level-$(i+1)$ such that after replacing the $M$-edge incident on $v$ with $vu$ in $M$, the vertex $u$ under the new edge set $M$ is not a $B'_1$-vertex.

The operation of \emph{eliminating a redundant vertex} $v$ is to update $M$ by replacing the $M$-edge incident on $v$ with $vu$, where $u$ is the vertex defined in the above definition of redundant vertices.

Our algorithm ${\tt crown}(G,A,p)$ is simple, the only main step of which is to eliminate redundant vertices.
Once there are no redundant vertices left, we let $C$ be the set of all vertices reachable from $B_1$ via strong $M$-alternating paths, including $B_1$, and let $I$ be the leaf neighbors of $C$ that participate in $M$.
The whole algorithm is described in Figure~\ref{crown}.

\begin{figure*}

\rule{\linewidth}{0.4mm}

\textbf{Input}: A graph $G=(V,E)$, a vertex subset $B\subseteq V$ and an integer $p\geq 1$ such that each connected component in $G[V\setminus B]$ has size at most $p$. \\
\textbf{Output}:  Two sets $I \subseteq V\setminus B$ and $C \subseteq B$ such that $(I,C,J=V\setminus(I\cup C))$ is a
$p$-weighted crown decomposition of $G$.
\begin{enumerate}
\item Let $A=V\setminus B$ and compute the auxiliary bipartite graph $H_G=(A',B',E')$ of $G$ with partition $(A,B)$.
\item For each vertex $v\in A'$, arbitrarily select an edge incident on $v$ and include the edge to $M$.
\item According to $M$, compute $A'_1, A'_2, A'_3, B'_1,B'_2$ and $B'_3$, and the hierarchical structure.
\item \textbf{If} \{there exists a redundant vertex\}, \\
\textbf{then} eliminate a redundant vertex and \textbf{goto} Step~3.
\item Let $C'$ be set of $B'$-vertices connected to a $B'_1$-vertex via a strong $M$-alternating path (also include the vertices in $B'_1$ themselves). Let $I'$ be the set of $A'$-vertices adjacent to a vertex in $C'$ via an edge in $M$.

\item Let $C$ be the set of vertices in $B$ in $G$ corresponding to the vertices in $C'$ in $H_G$ and $I$ be the set of
vertices in the components of $G[A]$ corresponding to the vertices in $I'$ in $H_G$.
\item \textbf{Return} ($I, C$).

\end{enumerate}

\rule{\linewidth}{0.4mm}
\caption{Algorithm ${\tt crown}(G,B,p)$}\label{crown}
\end{figure*}

\lem{redundant}{If there is no redundant vertex, then there is no strong $M$-alternating path from a $B'_1$-vertex to a $B'_3$-vertex.}
In fact, if there is a  strong $M$-alternating path $P$ from a $B'_1$-vertex to a $B'_3$-vertex $u$, we can prove that the vertex $v$ adjacent to $u$ in $P$ is a redundant vertex. Full proofs for the following lemmas can be found in Appendix~\ref{ap_proof}.

\lem{crown0}{Algorithm ${\tt crown}(G=(V,E),B,p)$ runs in $O(|V|^2|E|)$ time.
}

\lem{crown1}{Algorithm ${\tt crown}(G,B,p)$ returns two sets $I \subseteq V\setminus B$ and $C \subseteq B$ such that $(I,C,J=V(G)\setminus(I\cup C))$ is a
$p$-weighted crown decomposition of $G$.
}
\lem{crown2}{
The two sets $I$ and $C$ returned by ${\tt crown}(G,B,p)$ satisfy the following size condition:
\eqn{size1}{|V\setminus (B\cup I)|\leq (2p-1)|B\setminus C|.}
}

\refl{crown0}, \refl{crown1} and \refl{crown2} directly imply \refl{wexpansion}. We can get a kernel of
$2p(p+1)k$ vertices for  \textsc{$p$-Size Separator} by \refl{wexpansion}. Our algorithm first selects a maximal packing of
connected $(p+1)$-subgraphs and let $B$ denote the vertex set of it. Then $|B|\leq (p+1)\gamma_p (G)$.
If the size of $A=V\setminus B$ is greater than $(2p-1)|B|$, we just
reduce the graph by removing the two sets returned by ${\tt crown}(G,B,p)$. So we can bound the number of vertices in the graph
by $(2p-1)|B|+|B|\leq 2p(p+1)\gamma_p (G)$.

\lem{kernel1}{There is an algorithm that runs in $O(mn^2)$ time to return a kernel of
at most $2p(p+1)k$ vertices for \textsc{$p$-Size separator}.}

\section{The proof of \reft{our-thm} and an improved kernel} \label{kernel2}
\refl{wexpansion} in Section~\ref{kernel1} can be used to reduce the input graph with a vertex partition $(A=V\setminus B, B)$, where
$B$ is taken as the vertex set of a packing of connected $(p+1)$-subgraphs.
The size of $B$ is at most $(p+1)\gamma_p (G)$. \refl{wexpansion} guarantees that the size of $A$ is $O(p|B|)$.
Then we get a bound of $O(p^2\gamma_p (G))$ on the graph size for our problem.
In this section, we will improve the bound from $O(p^2\gamma_p (G))$ to $O(p\gamma_p (G))$ by proving \reft{our-thm}.
Our initial idea to get the improvement is, not to apply  \refl{wexpansion} on the whole graph $G$, but to find two subsets $A_1\subseteq A$ and $B_1\subseteq B$
and apply \refl{wexpansion}
only to the subgraph $G[A_1\cup B_1]$. If $|A\setminus A_1|+|B\setminus B_1|\leq p\gamma_p (G)$ and
$|B_1|\leq \gamma_p (G)$, then we may be able to bound the size of $A_1$ by $O(p|B_1|)=O(p \gamma_p (G))$ and
the size of the whole instance by $O(p\gamma_p (G))$.
The crucial parts of this method are to find the partition $(A,B)$ and the two subsets $A_1$ and $B_1$.
In fact, our algorithm does not really find the two subsets $A_1$ and $B_1$. We will still apply
\refl{wexpansion} on the whole graph $G$ with partition $(A,B)$. It is necessary to find $I$ and $C$ only in $A_1$ and $B_1$
for two subsets $A_1$ and $B_1$ satisfying the above properties.
Before introducing the main algorithm, we first derive a property and sub-algorithm that will be used later.

\subsection{The local adjustment property}
We introduce a property, which will be used to find the partition
$(A,B)$ of the input graph. Recall that a vertex set in a graph is a \emph{$p$-size separator}
if after deleting it each connected component of the remaining graph has at most $p$ vertices.
For a $p$-size separator containing only one vertex, the unique vertex in it is also called a \emph{$p$-size separator vertex}.

\thm{p_1}{
\emph{\textbf{[Local Adjustment Property]}}
Let $p\geq 1$ be an integer and $G=(V,E)$ be a connected graph of more than $3p$ vertices.
If $G$ has no $p$-size separator vertex, then there are two vertex-disjoint subsets $V_1,V_2\subseteq V$ such that
$|V_i|\geq p+1$ and the induced graph $G[V_i]$ is a connected graph for $i=1,2$.
Furthermore, the two subsets $V_1$ and $V_2$ can be computed in $O(p|E|)$ time.}

To prove \reft{p_1}, we show that the following algorithm ${\tt connect}(G,p)$ correctly computes the vertex subsets $V_1$ and $V_2$.

\begin{figure*}

\rule{\linewidth}{0.4mm}

\textbf{Input}: A connected graph $G=(V,E)$ of more than $3p$ vertices having no $p$-size separator vertex, and an integer $p>0$. \\
\textbf{Output}: Two vertex-disjoint subsets $V_1,V_2\subseteq V$ such that
$|V_i|\geq p+1$ and the induced graph $G[V_i]$ is a connected graph for $i=1,2$.
\begin{enumerate}
\item Initially $V_1 \leftarrow \emptyset$ and $V_2 \leftarrow V$.
\item \textbf{If} \{ $|V_1|\leq p$\}, \textbf{do}
\begin{enumerate}
  \item[2.1] Select a vertex $v\in N(V_1)$ (here we let $N(\emptyset)=V$) such that $V_1\cup \{v\}$ is not a $p$-size separator and let
$V_1 \leftarrow V_1\cup \{v\}$.
  \item[2.2] Let $U$ be a maximum connected component in $G\setminus V_1$,  $V_2 \leftarrow V(U)$ and $V_1 \leftarrow V\setminus V_2$.
\end{enumerate}
\item \textbf{Else} \textbf{return} $(V_1,V_2)$.
\end{enumerate}

\rule{\linewidth}{0.4mm}
\caption{Algorithm ${\tt connect}(G,p)$}\label{connect}
\end{figure*}

First, we show that after each iteration of Step~2,  $G[V_1]$ and $G[V_2]$ are two connected graphs.
Since $V_2$ is taken as the vertex set of a connected component of $G\setminus V_1$ in Step 2.2,
we know that $G[V_2]$ is a connected graph after executing Step 2.2.
Assume that $G[V_1]$ is a connected graph before executing an iteration of Step~2, which holds before
the first execution of Step~2 since we regard an empty graph as connected.
We only need to show that $G[V_1]$ is still a connected graph after executing an iteration of Step~2.
Let $V'_1=V_1$ and $V'_2=V_2$ at the time before executing an iteration of Step~2. Then both of $G[V'_1]$ and $G[V'_2]$ are connected graphs.
After executing Step 2.1, $G[V_1=V'_1\cup\{v\}]$ is still connected since we select $v$ from $N(V'_1)$.
Since $G[V'_2]$ is a connected graph, each connected component of $G[V'_2\setminus \{v\}]$ has some vertex adjacent to
$v$. Then for any connected component $U_0$ of $G[V'_2\setminus \{v\}]$, the graph $G[V'_2\setminus V(U_0)]$ is still
connected. After Step 2.2, $V_1=V\setminus V(U)= (V'_1\cup \{v\})\cup (V'_2\setminus V(U))$ for a connected component
$U$ of $G[V'_2\setminus \{v\}]$. Thus, $G[V_1]$ is still a connected graph.

Second, we show that the vertex $v$ in Step 2.1 always exists, which is important for the correctness of the algorithm.
It is clear that $(V_1,V_2)$ is a partition of the vertex set $V$ before or after an execution of Step~2.
We still let $V'_1=V_1$ and $V'_2=V_2$ at the time before executing an iteration of Step~2. Then both of $G[V'_1]$ and $G[V'_2]$ are connected graphs by the above analysis.
We know that $N(V'_1)\neq \emptyset$ since $G$ is a connected graph. Let $v_0$ be an arbitrary vertex in $N(V'_1)$.
Assume that $V'_1\cup \{v_0\}$ becomes a $p$-size separator of $G$. Let $\mathcal{U}=\{U_i\}$ be the set of connected components
of $G\setminus (V'_1\cup \{v_0\})$. Then each $U_i$ contains at most $p$ vertices.
Since $v_0$ is not a $p$-size separator vertex of $G$, we know that there is a vertex $v'_0$ in $U_{i_0}$ for some $U_{i_0} \in \mathcal{U}$ such that
$v'_0$ is adjacent to a vertex in $V'_1$, i.e., $v'_0 \in N(V'_1)$.
Next, we show that $V'_1\cup \{v'_0\}$ is impossible to be a $p$-size separator again. Let
$\mathcal{W}=\{W_i\}$ be the set of connected components
of $G\setminus (V'_1\cup \{v'_0\})$. Note that $v'_0\in U_{i_0}$ and then $(\cup_{i\neq i_0}U_{i})\cup\{v_0\} \subseteq W_{i_1}$
for some $W_{i_1} \in \mathcal{W}$.
If $V'_1\cup \{v'_0\}$ is a $p$-size separator, then $|(\cup_{i\neq i_0}U_{i})\cup\{v_0\}|\leq |W_{i_0}|\leq p$.
Therefore, $|V|=|V'_1|+ |V'_2|\leq p+ |(\cup_{i\neq i_0}U_{i})\cup\{v_0\}\cup U_{i_0}|
\leq p+ p+p=3p$, a contradiction with the graph $G$ having more than $3p$ vertices.
So $V'_1\cup \{v'_0\}$ is impossible to be a $p$-size separator. The vertex $v$ in Step 2.1 always exists and it can be
found  in linear time by checking at most two vertices in $N(V'_1)$.

Finally, we are ready to prove the correctness of the whole algorithm and analyze the running time.
After Step~2, we have that $|V_1|\geq p+1$. Note that in Step 2.1, the vertex $v$ is selected such that $V_1\cup \{v\}$ is not a $p$-size separator vertex and
in Step 2.2, $U$ is selected as a component of maximum size. So we know that $|V_2|=|V(U)|\geq p+1$.
We have proved that after Step~2,  $G[V_1]$ and $G[V_2]$ are two connected graphs.
Then the two sets $V_1$ and $V_2$ returned in Step 3 satisfy the requirement.
Next, we only need to show that the algorithm always stops in $O(p|E|)$ time.
We have shown that the vertex $v$ in Step 2.1 can be found in $O(|E|)$ time. Then Step 2 can be executed in $O(|E|)$ time.
No vertex in $V_1$ will be moved out of $V_1$ and in each iteration of Step 2 at least one new vertex in included to $V_1$.
So Step~2 can be executed for at most $p+1$ iterations and the algorithm always stops and runs in $O(p|E|)$ time.

In fact, the bound $3p+1$ on the number of vertices in \reft{p_1} is tight.
For if a graph has only $3p$ vertices but has no $p$-size separator vertex, it may not have two vertex subsets $V_1$ and $V_2$
satisfying the conditions in \reft{p_1}. A graph of a triangle is an example for $p=1$.

\subsection{The main algorithm}
Our algorithm is still based on a vertex partition $(A,B=V\setminus A)$ of the graph.
For a subset $B^*\subseteq B$, a component of $G[A]$ is called a \emph{$B^*$-attached component} if the component has a vertex adjacent to some vertex in $B^*$.
We use $G(B^*)$ to denote the graph induced by $B^*$ together with the vertices of all $B^*$-attached components.
We call $G(B^*)$ the \emph{$B^*$-associate subgraph}.

In our algorithm, the partition $(A,B)$ is not fixed. However, through the algorithm, $B$ always satisfies the \emph{base-properties}:
\begin{enumerate}
\item[(\textbf{Q1})] $B$ is a $p$-size separator, i.e.,
each connected component of $G[A]$ has at most $p$ vertices.
\item[(\textbf{Q2})] $B$ is the union of several pairwise disjoint vertex sets, called \emph{bases}, each of which is either a set of a single vertex or the vertex set of a $(p+1)$-subgraph.
A base of a single vertex is called a \emph{single-base} and a base of $p+1$ vertices is called a \emph{group-base}.
\item[(\textbf{Q3})] The number of bases in $B$ does not decrease in the algorithm and it is at most $\gamma_p (G)$.
\item[(\textbf{Q4})] Let $H_G=(A',B',E')$ be the auxiliary bipartite graph of $G$ with partition $(A,B)$. Let $B'_s\subseteq B'$ be the subset of vertices in $B'$ corresponding to
the set of vertices in single-bases in $B$. There is a full $(4p,\infty)$-star packing from $B'_s$ to $A'$.
\end{enumerate}

Figure~\ref{figt4} in Appendix~\ref{fig} gives an illustration for such partition $(A,B)$ with the base-properties.
Initially $B$ contains only group-bases. For this case, $B$ is the vertex set of a maximal $(p+1)$-subgraph packing and  the base-properties clearly hold.
Then the algorithm tries to increase either the number of bases in $B$ by replacing a group-base with two group-bases or the number of single-bases in $B$ by replacing a group-base with a single-base.
Let $S$ be a group-base in the current $B$ (may have always contained some single-bases). We consider the $S$-associate subgraph $G(S)$.

\textbf{Case 1.} $G(S)$ has at least $3p+1$ vertices and has no $p$-size separator vertex: we call the group-base $S$ \emph{extendable}.
For this case, $G(S)$ satisfies the conditions of the input graph of  ${\tt connect}$.
We can get two connected components $S_{n1}$ and $S_{n2}$ of $p+1$ vertices in $G(S)$ by using \reft{p_1} and the algorithm ${\tt connect}$. Then we replace $S$ with two group-bases $S_{n1}$ and $S_{n2}$ in $B$. Furthermore, we can extend each single-base to a group-base in $B$ (keeping all bases pairwise vertex-disjoint).
The reason why we need to replace single-bases with group-bases is to show that the total number of bases in $B$  is at most $\gamma_p (G)$ in (\textbf{Q3}).
We also need (\textbf{Q4}), because it is crucial for us to  extend single-bases to group-bases keeping all bases pairwise vertex-disjoint.
In our algorithm, when we find an extendable group-base, we will increase the number of bases in $B$ by one
 according to this. This operation is called `Extension Operation'. We will introduce the details of it later.
Note that Extension Operation can be executed for at most $\gamma_p (G)$ times.

\textbf{Case 2.} $G(S)$ has a $p$-size separator vertex $v$ or has at most $3p$ vertices: we use $\bar{B}$ to denote the set after replacing $S$ with a single-base $v$ in $B$.
We call the group-base $S$ \emph{changeable} if $\bar{B}$ fulfills (\textbf{Q4}). When we find a changeable group-base $S$, we replace $S$ with the vertex $v$ in $B$.
It is easy to see that after this operation, $B$ still holds the base-properties.

Our algorithm will iteratively deal with extendable and changeable group-bases found by the algorithm.
Finally, the remaining group-bases will not cause troubles in our analysis.
We construct a graph $G^*$ by merging each group-base in $G$ into a single vertex, called \emph{group-vertex}.
Let $B^*$ be the set after replacing each group-base with a group-vertex in $B$. 
We  invoke  ${\tt crown}(G^*,B^*,p)$. Let $(I,C)={\tt crown}(G^*,B^*,p)$.
We will guarantee that: the set $C\subseteq B^*$ returned by  ${\tt crown}(G^*,B^*,p)$ does not contain any group-vertex, $|A\setminus C|=|A^*\setminus C|=O(pk)$, and
$|B\setminus C|=O(pk)$.
Then we can reduce the instance by removing $I\cup C$ in $G$ by \refl{wexpansion}. The remaining part of the graph has only $O(pk)$ vertices.
The main steps of the algorithm to compute $I$ and $C$ are presented in Figure~\ref{whole}. Next, we give the details of some steps in the algorithm and proofs.

\begin{figure*}[h]

\rule{\linewidth}{0.4mm}

\textbf{Input}: A connected graph $G=(V,E)$ and an integer $p>0$. \\
\textbf{Output}: Two vertex subsets $I$ and $C$ such that $(I,C, J=V\setminus (I\cup C))$ a $p$-weighted crown decomposition of $G$ satisfying the size condition in \reft{our-thm}.
\begin{enumerate}
\item Compute a maximal packing $\mathcal{S}$ of connected $(p+1)$-subgraphs.
\item let $B$ be the set of vertices appearing in $\mathcal{S}$ and let $A=V\setminus B$, where the vertex set of each $(p+1)$-subgraph in $\mathcal{S}$ is a group-base in $B$.
\item \textbf{If} \{there exits an extendable group-base $S$\}, \textbf{do}
\begin{enumerate}
\item [3.1]increase the number of bases in $B$ according to Extension Operation: first replace $S$ with two new
group-bases and then extend each single-base to a group-base;
\item[3.2]update $\mathcal{S}$ by letting it be the packing of connected $(p+1)$-subgraphs corresponding to the group-bases in $B$;
\item[3.3] \textbf{goto} Step 2.
\end{enumerate}
\item construct $G^*$ from $G$ by merging each group-base into a group-vertex, and let $B^*$ be the set obtained by replacing each group-base with a group-vertex in $B$. Let  $(I,C)={\tt crown}(G^*,B^*,p'=4p)$.
    ~~~~~~~// We also keep original $G$ and $B$.
\item \textbf{If} \{A vertex $u\in C$ is a group-vertex\}, \textbf{do}~~~~~~// Then $u$ should be changeable
\begin{enumerate}
\item [5.1] update $B$ by replacing $S$ ($S$ is the group-base corresponding to the group-vertex $u$) with a single-base $v$ in $B$ and let $A=V\setminus B$, where $v$ is a $p$-size separator vertex of $G(S)$;
\item[5.2] \textbf{goto} Step 3.
\end{enumerate}

\item \textbf{return} $(I,C)$ ( Now $C$ does not contain any group-vertex and then vertex sets $I$ and $C$ returned by ${\tt crown}(G^*,B^*,p'=4p)$ are also two vertex sets in the original graph $G$).
\end{enumerate}

\rule{\linewidth}{0.4mm}
\caption{Algorithm ${\tt SCC}(G,p)$}\label{whole}
\end{figure*}

\textbf{Step 3.}
Let $S$ be an extendable group-base in Step~3. Then $G(S)$ is a connected graph of at least $3p+1$ vertices and it has no $p$-set separator vertex.
By invoking ${\tt connect}(G(S),p)$, we can find two connected subgraphs $G[V_1]$ and $G[V_2]$ of at least $p+1$ vertices, where $V_1\cup V_2= V(G(S))$.
Note that each $S$-attached component has at most $p$ vertices and then each of $V_1$ and $V_2$ must contain some vertices in $S$.
We look at $G[V_i\setminus S]$ ($i=1,2$), each connected component of it is a $S$-attached component or a partial of a $S$-attached component. After removing some of such components from $G[V_i]$ the remaining graph is still connected.
So we iteratively remove a connected component of $G[V_i\setminus S]$ from $G[V_i]$ until $G[V_i]$ has less than
$2p+1$ vertices. We use $V'_i$ to denote the current set $V_i$. Now $G[V'_i]$ is still connected and has at least $p+1$ vertices since each connected component of $G[V_i\setminus S]$
has at most $p$ vertices. We select a set $V''_i$ of exactly $p+1$ vertices from $V'_i$ such that $G[V''_i]$ is
a connected graph. The two $(p+1)$-subgraphs $G[V''_1]$ and $G[V''_2]$ are what we are looking for.
We update $B$ by replacing $S$ with two group-bases $S_{n1}=V''_1$ and $S_{n2}=V''_2$ in $B$.
Next, we show that each single-base in $B$ can be extended to a group-base keeping all vertices in $B$ different.
By (\textbf{Q4}), we know that there is  a full $(4p,\infty)$-star packing from $B'_s$ to $A'$ (before replacing
$S$ with $S_{n1}$ and $S_{n2}$). So each vertex $v\in B'_s$ (a vertex in a single-base) has the some `exclusive'
$\{v\}$-attached components (which are corresponding to the leaves of the star centered at $v$ in the full $(4p,\infty)$-star packing). The number of vertices in exclusive
$\{v\}$-attached components is at least $4p$. Some exclusive $\{v\}$-attached components may be included to $V_1\cup V'_2$. However, $|V_1\cup V_2\setminus S|\leq 2p+2p-(p+1)=3p-1$. The number of vertices in the
exclusive $\{v\}$-attached components not containing any vertices in $V_1\cup V_2$ is at least $4p-(3p-1)=p+1$.
These vertices together with $v$ will form a connected component of at least $p+1$ vertices.
We extend a single-base $\{v\}$ to a group-base by using these vertices. Note that each  single-base $v$ only uses
exclusive $\{v\}$-attached components to extend to a group-base. So all the group-bases will be vertex-disjoint.
The above operation is called \emph{Extension Operation}.

Next, we analyze the running time bound of Step~3. In fact, the full $(4p,\infty)$-star packing used in this step is computed in Step~4. So in this step, we mainly use $O(pm)$ time to invoke ${\tt connect}(G(S),p)$. All the other operations can be implemented in linear time. So Step~3 uses $O(pm)$ time.

\textbf{Step 5.} We mainly prove the claim in this step: if a vertex $u\in C$ is a group-vertex, then $u$ is
a changeable group-vertex. Let $S$ be the group-base in $B$ corresponding to the group-vertex $u\in B^*$.
Step~3 has been executed and then $S$ is not extendable.
In fact, the $S$-associate subgraph $G(S)$ has at least $p'+(p+1)=5p+1>3p$ vertices since $u\in C$ and there is a
$(p',\infty)$-star from $u$ to vertices in $I$. So we know that $G(S)$ has a $p$-size separator vertex $v$.

To prove that $S$ is changeable, we need to consider (\textbf{Q4}). Let $B^*_s\subseteq B^*$ be the set of vertices corresponding to the vertices in single-bases in $B$. There is a full $(p'=4p,\infty)$-star packing from $(C\cap B^*_s)\cup \{u\}$ to $I$. Note that there are no edges between $I$ and $B^*\setminus C$. Then there is
a full $(4p,\infty)$-star packing from $B^*_s\setminus C$ to $A^*\setminus I=V(G^*)\setminus B^* \setminus I$ since
(\textbf{Q4}) holds on $B$.  Then there is a full $(4p,\infty)$-star packing from $B^*_s\cup\{u\}$ to $A^*$,
which implies that (\textbf{Q4}) holds on $\bar{B}$, where $\bar{B}$ is obtained by replacing $S$ with $v$.
Thus, $u$ is a  changeable group-vertex.

In Step~5, after replacing a changeable group-base $S$ with a single-base $v$, $B$ still has the base-properties:
It is easy to see that (\textbf{Q2}) and (\textbf{Q3}) trivially hold; The definition of changeable group-bases directly implies  (\textbf{Q1}) and (\textbf{Q4}).

Steps 4-5 mainly use $O(n^2m)$ time to invoke the algorithm ${\tt crown}(G^*,B^*,p')$. Other operations can be executed in linear time.

\medskip

Based on the above analysis for Steps 3-5, we are ready to prove the main results:

\lem{result1}{Algorithm ${\tt SCC}(G,p)$ runs in $O(\gamma_p (G)^2n^2m)$ time and the two sets $I$ and $C$ returned by ${\tt SCC}(G,p)$ make $(I,C, J=V\setminus (I\cup C))$ a $p$-weighted crown decomposition of $G$.}

A proof of \refl{result1} can be found in Appendix~\ref{ap_proof}.

\lem{result2}{The two sets $I$ and $C$ returned by ${\tt SCC}(G,p)$ satisfy the size condition in \reft{our-thm}, i.e.,
$|V(G) \setminus (C \cup I)|\leq 9p |K|$, for any $p$-size separator $K$ of the induced subgraph $G[V \setminus (C \cup I)]$.
}
\pf{We consider the graph $G^*$ constructed in Step~4 and the original graph $G$. It is easy to see that
$$ |B^*\setminus C|\leq |K|.$$
By \refl{crown2}, we know
that $|V(G^*)\setminus (B^*\cup I)|\leq (2p'-1)|B^*\setminus C|$. Note that $V(G)\setminus (B\cup I)=V(G^*)\setminus (B^*\cup I)$.
Then we have that
 $$|V(G)\setminus (B\cup I)|\leq (2p'-1)|K|=(8p-1)|K|.$$

 Let $x$ be the number of group-vertices in $B^*$.
We know that $x\leq |B^*\setminus C|$ and $|B\setminus C|=|B^*\setminus C|+xp$.
Thus,
$$|B\setminus C|=(p+1)|B^*\setminus C|\leq (p+1)|K|.$$
By the above two inequalities, we get that
$$|V(G) \setminus (C \cup I)|=|V(G)\setminus (B\cup I)|+|B\setminus C|\leq 9p|K|.$$
}

\reft{our-thm} directly follows from \refl{result1} and \refl{result2}, which also implies:

\crl{kernel}{\textsc{$p$-Size separator} admits a kernel of $9pk$ vertices.}

\section{Concluding remarks} \label{conclude}

In this paper, we first introduce a weighted crown decomposition and a
weighted expansion lemma, and obtain an $O(p^2k)$ vertex kernel for \textsc{$p$-Size separator} by using them.
The weighted expansion lemma is not a simple extension of the original expansion lemma. We need to use a `hierarchical structure' to prove it.
Then we further improve the kernel bound from $O(p^2k)$ to $O(pk)$ by proving \reft{our-thm}.
This theorem can be regarded as a generalization of the Nemhauser and Trotter's
local optimization theorem for \textsc{vertex Cover} and it is proved
based on extremal combinatorial arguments. The improvement is obtained by using a `local adjustment property' and some
other techniques. These techniques are possible to be used to improve kernel sizes for more problems, especially
problems with kernelization algorithms based on techniques similar to the expansion lemma or crown decompositions.



\newpage
\appendix

\centerline{\bf\large Appendix}

\section{Figures}\label{fig}

\begin{figure}[!h]
\begin{center}
\includegraphics[width=0.8\textwidth]{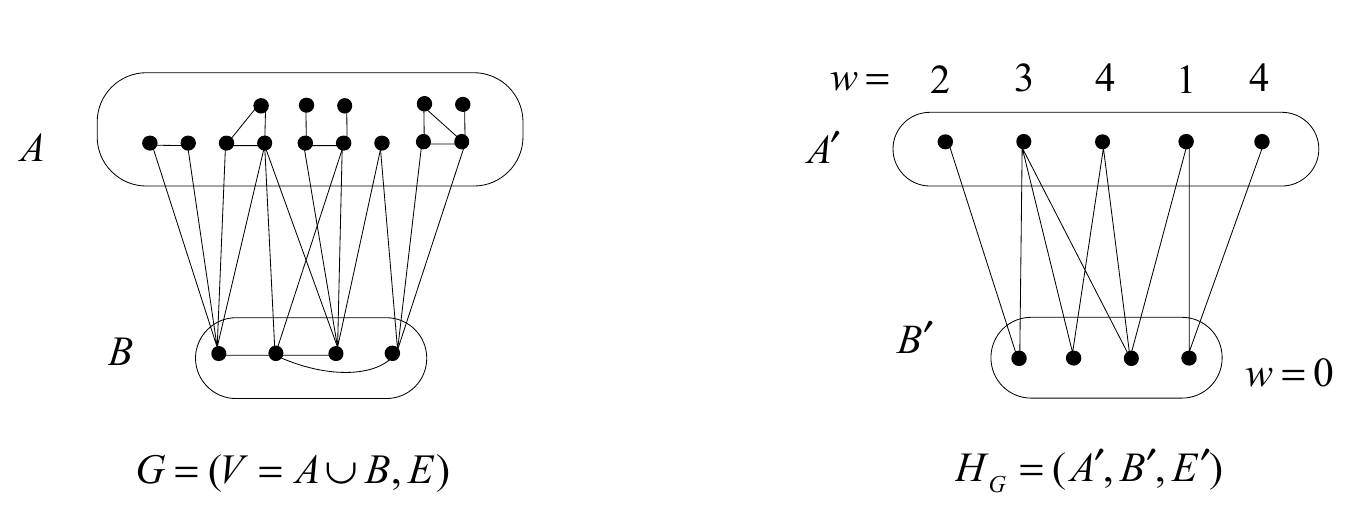}
\end{center}
\caption{The graph $G$ and the auxiliary bipartite graph $H_G$ of $G$ with partition $(A,B)$}\label{figt1}
\end{figure}

\begin{figure}[!h]
\begin{center}
\includegraphics[width=0.9\textwidth]{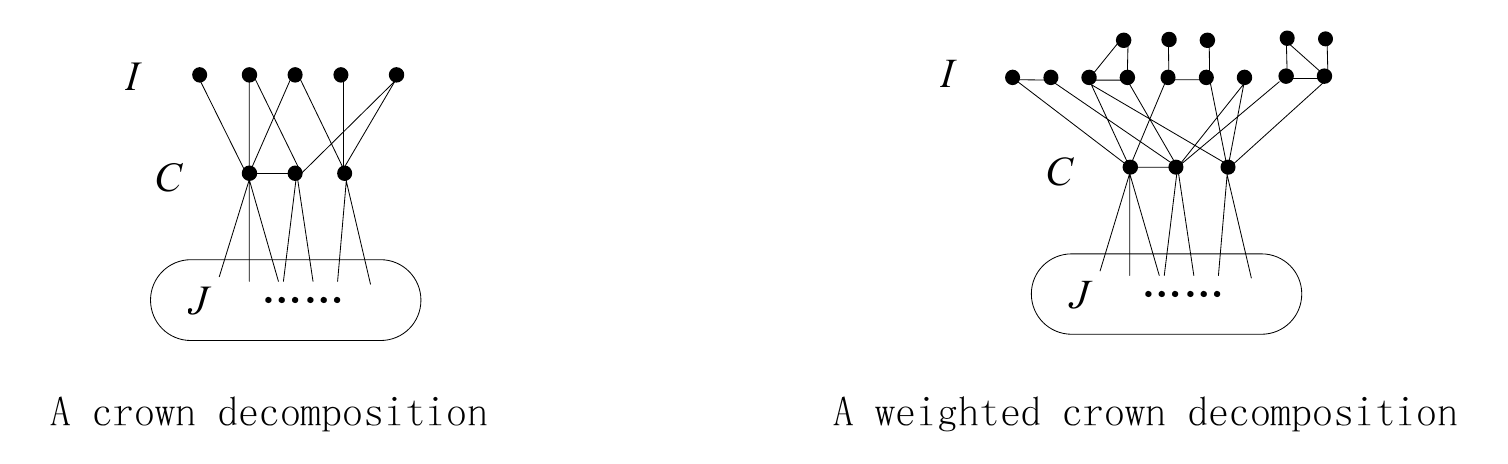}
\end{center}
\caption{Crown decompositions and weighted crown decompositions}\label{figt2}
\end{figure}

\begin{figure}[!h]
\begin{center}
\includegraphics[width=0.6\textwidth]{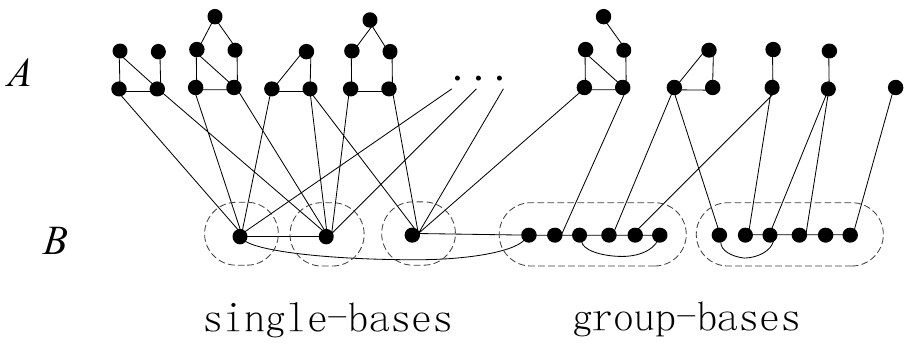}
\end{center}
\caption{A partition $(A,B)$ holding the base-properties}\label{figt4}
\end{figure}

\section{Some Proofs}\label{ap_proof}

\noindent \textbf{\refl{wcrown}}
\emph{{Let $(I,C,J)$ be a $p$-weighted crown decomposition of a graph $G=(V,E)$. Then $(I,C)$ satisfies the local optimality condition in \reft{our-thm}, i.e.,
$K \cup C$ is a minimum $p$-size separator of $G$ for any minimum $p$-size separator $K$ of the induced subgraph $G[J]$.
}}

\medskip
\pf{ It is easy to see that $K \cup C$ is a $p$-size separator of $G$: by \refd{weight_dc} we know that each vertex in $I$ is in a component of size at most $p$ in the remaining graph after deleting $C$; and each vertex in $J\setminus K$ is in a component of size at most $p$ in the remaining graph after deleting $C\cup K$ since $K$ is a $p$-size separator of subgraph $G[J]=G[V\setminus(I \cup C)]$.

Next, we show that $K \cup C$ is also a minimum $p$-size separator of $G$.
Let $D$ be an arbitrary minimum $p$-size separator of $G$.
Let $D_1=D\cap (I\cup C)$ and $D_2=D\cap J$.
Since there are $|C|$ vertex-disjoint connected subgraphs of $G[I\cup C]$ each of which contains at least $p+1$ vertices,
each $p$-size separator contains at least $|C|$ vertices in $I\cup C$. Then we have that
$$|D_1|\geq |C|.$$

Set $D_2$ is a $p$-size separator of $G[V\setminus D_1]$ and
set $K$ is a minimum $p$-size separator of $G[V\setminus (I \cup C)]$.
Note that $D_1 \subseteq I\cup C$ and then $G[V \setminus (I \cup C)]$ is an induced subgraph of $G[V\setminus D_1]$.
The size of a minimum $p$-size separator of $G[V\setminus D_1]$ will not be smaller than
the size of a minimum $p$-size separator of $G[V\setminus (I \cup C)]$. Thus,
$$|D_2|\geq |K|.$$
Therefore,  $|K \cup C|=|K|+|C|\leq |D_1|+|D_2|=|D|$.
}

\medskip

\noindent \textbf{\refl{redundant}}
\emph{{If there is no redundant vertex, then there is no strong $M$-alternating path from a $B'_1$-vertex to a $B'_3$-vertex.}}

\medskip
\pf{Assume that there is a strong $M$-alternating path from a $B'_1$-vertex to a $B'_3$-vertex $u$.
Then there is a pure $M$-alternating path $P$ from a $B'_1$-vertex to $u$ by properties (\textbf{P2}) and (\textbf{P3}).
Assume that $u$ is in level-$i$ in the hierarchical structure.
Then $i$ is an even number since $u$ is a $B'$-vertex by (\textbf{P1}).
We show that the vertex $v$ adjacent to $u$ in $P$ is a redundant vertex.
Since $u$ is a $B'$-vertex and $P$ is a strong $M$-alternating path from a $B'_1$-vertex to $u$,
we know that $vu$ is an edge not in $M$ and then $v$ is in level-$(i-1)$ in the hierarchical structure,
where $i-1$ is an odd number.
The total weight of the vertices adjacent to $u$ via an $M$-edge is less than $p$ since $u\in B'_3$ and $w(v)\leq p$ since $v\in A'$. Then after adding $uv$ to $M$ (also deleting the original $M$-edge incident on $v$), the
total weight of the vertices adjacent to $u$ via an $M$-edge is less than $2p$. So $v$ is a redundant vertex and then
the lemma holds.}

\medskip
\noindent \textbf{\refl{crown0}}
\emph{{Algorithm ${\tt crown}(G=(V,E),B,p)$ runs in $O(|V|^2|E|)$ time.}}

\medskip
\pf{ Step~5 can be executed in linear time via a BFS. It is easy to see that each of other steps takes only linear time. Then we only need to analyze how many loops of Steps 3-4 will be executed.

Let $v$ be a redundant vertex in level-$i$. Then $v$ is adjacent to a level-$(i-1)$ vertex $v_0$ via an $M$-edge $v_0v$.
In the operation of eliminating $v$ in Step~4, we will replace the edge $v_0v$ with another edge $vv^*$ in $M$,
where $v^*$ is a $B'$-vertex in level-$(i+1)$.
After this operation, no new $B'_1$-vertex is created and no vertex will be move to a `higher' level with a smaller index.
At least $v$ will appear in a level with index greater than $i$ or not appear in the hierarchical structure any more.
So after this operation some vertices (at least one vertex) either move to a level with larger index or
disappear from the hierarchical structure. So Step~4 can be executed for at most $|V|^2$ time.
Then the algorithm always stops and runs in $O(|V|^2|E|)$ time.
}

\medskip
\noindent \textbf{\refl{crown1}}
\emph{{Algorithm ${\tt crown}(G,B,p)$ returns two sets $I \subseteq V\setminus B$ and $C \subseteq B$ such that $(I,C,J=V(G)\setminus(I\cup C))$ is a
$p$-weighted crown decomposition of $G$.
}}

\medskip
\pf{We have shown in \refl{crown0} that ${\tt crown}(G,B,p)$ always stops. To prove the correctness, we check the three conditions in the definition of $p$-weighted crown decompositions.

According to the definition of $C'$ and \refl{redundant}, we know that $C'$ is a subset of $B'_1\cup B'_2$.
Therefore, there is a full $(p,\infty)$-star packing from $C'$ to $I'$, which is formed by the edges in $M$. Condition (3) in \refd{weight_dc} holds.

Condition (2) trivially holds since $I$ is just a subset of $A=V(G)\setminus B$ and the input of ${\tt crown}(G,B,p)$ requires that each component of $G[A]$ has at most $p$ vertices.

Next we consider condition (1). We first show that there is no edge between $I'$ and $B'\setminus C'$. Assume to the contrary that there is an edge $uv$ between $I'$ and $B'\setminus C'$,
where $u \in I'$ and $v \in B'\setminus C'$.
We show that there is a strong $M$-alternating path $P$ from a $B'_1$-vertex to $v$.
Assume that $uv'\in M$. Then $v'$ is a vertex in $C'$ since $u$ is in $I'$.
There is a strong $M$-alternating path $P'$ from a $B'_1$-vertex $v^*$ to $v'$ according to the definition of $C'$. If $P'$ passes $u$, then let $P$ be the subpath of $P'$ from $v^*$ to $u$ adding an edge $uv$. Otherwise
we let $P$ be the path adding $v'uv$ to the end of $P'$.
We can see that $P$ is still a strong $M$-alternating path from a $B'_1$-vertex to $v$, which together with \refl{redundant} imply a contradiction that there is a redundant vertex.

Therefore, there is no edge between $I'$ and $B'\setminus C'$ in $H_G$ and
then there is no edge between $I$ and $B\setminus C$ in $G$.
It is clear that there is no edge between $I$ and $A\setminus I$ in $G$. So there is no edge between $I$ and $J=(A\setminus I)\cup (B\setminus C)$ in $G$. Condition (1) holds.

 All the three conditions in \refd{weight_dc} hold and then $I$ and $C$ make $(I,C,J=V(G)\setminus(I\cup C))$ a
$p$-weighted crown decomposition.      }

\medskip
\noindent \textbf{\refl{crown2}}
\emph{{
The two sets $I$ and $C$ returned by ${\tt crown}(G,B,p)$ satisfy the following size condition:
\eqn{size1}{|V\setminus (B\cup I)|\leq (2p-1)|B\setminus C|.}
}}

\medskip
\pf{According to the definition of $C'$ we know that $B'\setminus C'$ does not contain any vertex in $B'_1$.
Then the weight of vertices in $A'\setminus I'$ is bounded by $(2p-1)|B'\setminus C'|$, which implies \refe{size1}.
}

\medskip

\noindent\textbf{\refl{result1}}
\emph{{Algorithm ${\tt SCC}(G,p)$ runs in $O(\gamma_p (G)^2n^2m)$ time and the two sets $I$ and $C$ returned by ${\tt SCC}(G,p)$ make $(I,C, J=V\setminus (I\cup C))$ a $p$-weighted crown decomposition of $G$.}}

\medskip
\pf{
First, we prove that two sets $I$ and $C$ returned by ${\tt SCC}(G,p)$ make $(I,C, J=V\setminus (I\cup C))$ a $p$-weighted crown decomposition of $G$ by checking the three conditions in the definition of
$p$-weighted crown decompositions.

In Step~6, no vertex in $C$ is a group-vertex and then each vertex in $I$ and $C$ is a vertex in the original graph $G$. By \refl{crown1}, the two sets $I$ and $C$ returned by ${\tt crown}$ hold Condition (3) in \refd{weight_dc}.
There are also no edge between $I$ and $(B\setminus C)\cup (A\setminus I)$ by \refl{crown1}.
The condition of (1) in \refd{weight_dc} holds. Condition (2) in \refd{weight_dc} holds because the set $B$ always holds
the base-properties (we have shown above that after one execution of Step~3 the set $\mathcal{S}$ is still
a maximal connected $(p+1)$-subgraph packing and after one execution of Step~5 the set $B$ still holds the base-properties).

Second, we consider the running time bound. We have analyzed above each execution of Step~3 takes $O(pm)$ time and
each execution of Step~4 takes $O(n^2m)$ time. We analyze how many iterations Steps~3-5 will be executed.
Step~3 can be executed for at most $\gamma_p (G)$ times. Between two adjacent executions of Step~3, Steps~4-5 can be
executed for at most $\gamma_p (G)$ times since each execution of Step~5 will increase the number of single-bases, which will not decrease except executing Step~3. Therefore, Steps~4-5 can be executed for at most $\gamma_p (G)^2$ times.
The whole algorithm runs in $O(\gamma_p (G)^2n^2m)$ time.
}

\end{document}